\documentclass[ALICE,manyauthors]{cernphprep}
\pdfoutput=1
\usepackage[comma,square,numbers,sort&compress]{natbib}
\usepackage{hyperref}
\usepackage{lineno}
\usepackage{xcolor}
\usepackage{hyperref}
\usepackage[normalem]{ulem}

\definecolor{dgreen}{cmyk}{1.,0.,1.,0.2}
\definecolor{orange}{cmyk}{0.,0.353,1.,0.}

\begin{document}%

\begin{titlepage}
\PHnumber{115}    
\PHdate{5 May}              
\PHyear{2016}
%

\title{Pseudorapidity dependence of the anisotropic flow of charged particles in Pb--Pb collisions at $\sqrt{s_{\text{NN}}}=2.76$ TeV}
\ShortTitle{Pseudorapidity dependence of the anisotropic flow}   

\Collaboration{ALICE Collaboration\thanks{See Appendix~\ref{app:collab} for the list of collaboration members}}
\ShortAuthor{ALICE Collaboration} 

\begin{abstract}
We present measurements of the elliptic ($\mathrm{v}_2$), triangular ($\mathrm{v}_3$) and quadrangular ($\mathrm{v}_4$) anisotropic azimuthal flow over a wide range of pseudorapidities ($-3.5< \eta < 5$). The measurements are performed with Pb--Pb collisions at $\sqrt{s_{\text{NN}}} = 2.76$ TeV using the ALICE detector at the Large Hadron Collider (LHC). The flow harmonics are obtained using two- and four-particle correlations from nine different centrality intervals covering central to peripheral collisions. We find that the shape of $\mathrm{v}_n(\eta)$ is largely independent of centrality for the flow harmonics $n=2-4$, however the higher harmonics fall off more steeply with increasing $|\eta|$. We assess the validity of extended longitudinal scaling of $\mathrm{v}_2$ by comparing to lower energy measurements, and find that the higher harmonic flow coefficients are proportional to the charged particle densities at larger pseudorapidities. Finally, we compare our measurements to both hydrodynamical and transport models, and find they both have challenges when it comes to describing our data.
\end{abstract}
\end{titlepage}
\setcounter{page}{2}

\section{Introduction}
The main goal of the heavy-ion physics program at the Large Hadron Collider (LHC) is to study the quark-gluon plasma (QGP), a deconfined state of matter existing at extreme temperatures and energy-densities. Experimental results from RHIC were the first to suggest that the QGP behaves as a nearly perfect fluid \cite{Arsene:2004fa,Back:2004je,Adams:2005dq,Adcox:2004mh}. A particularly important observable when characterizing the QGP is anisotropic azimuthal flow. The anisotropic flow develops from pressure gradients originating from the initial spatial geometry of a collision and is observed as a momentum anisotropy in the final-state particles. It is usually described by flow harmonics, which are defined as the Fourier coefficients:
\begin{equation}
\mathrm{v}_n = \left\langle \cos \left[ n(\varphi-\Psi_n) \right] \right\rangle,
\end{equation}
where $n$ is the order of the flow harmonic, $\varphi$ is the azimuthal angle and $\Psi_n$ is the symmetry plane angle of harmonic $n$. The first three Fourier coefficients, $\mathrm{v}_1$, $\mathrm{v}_2$, and $\mathrm{v}_3$ are known as directed, elliptic and triangular flow, respectively. The flow harmonics $\mathrm{v}_1$ to $\mathrm{v}_6$ have been studied extensively at RHIC \cite{Arsene:2004fa,Back:2004je,Adams:2005dq,Adcox:2004mh,Adams:2003zg, Adare:2011tg, Adamczyk:2013waa} and the LHC \cite{Aamodt:2010pa,ALICE:2011ab,ATLAS:2011ah,Chatrchyan:2012wg,ATLAS:2012at,Chatrchyan:2012ta,Aad:2013xma,Abelev:2013cva,Chatrchyan:2013kba,Abelev:2014pua}. The observed anisotropic flow is considered to be a strong indication of collectivity \cite{Ollitrault:1992bk} and is described well by relativistic hydrodynamics \cite{Luzum:2008cw}. 

Anisotropic flow studies at RHIC played a major role in establishing that the produced system is a strongly interacting quark-gluon plasma (sQGP) \cite{Arsene:2004fa,Back:2004je,Adams:2005dq,Adcox:2004mh} with a shear viscosity to entropy density ratio ($\eta/s$) close to the conjectured lower limit of $1/(4\pi)$ predicted by the AdS/CFT correspondence \cite{Kovtun:2004de}. The fact that higher order harmonics are increasingly suppressed by viscosity \cite{Alver:2010dn} makes it possible to use anisotropic flow measurements to estimate the $\eta/s$ of the produced system \cite{Luzum:2012wu,Gardim:2014tya}.

The pseudorapidity ($\eta$) dependence of the flow harmonics can play a key role in understanding the temperature dependence of $\eta/s$, something that can be determined using Quantum Chromodynamics (QCD) \cite{ Prakash:1993bt, Arnold:2003zc, Denicol:2015nhu}.  At forward rapidities, the average temperature drops which implies $\eta/s$ will also change. In addition, the lower temperatures at forward rapidities mean the system will spend less time in the QGP phase leading to the hadronic viscosity playing a greater role in affecting the flow harmonics \cite{Molnar:2014zha, Denicol:2015nhu}. Recently, it has been suggested that the symmetry plane angles may depend on $\eta$ \cite{Gardim:2012im,Jia:2014vja,Khachatryan:2015oea}. While this effect is not directly studied in this Letter, considering that the reference particles are taken from mid-rapidity, the measured values of anisotropy coefficients at forward rapidity will be suppressed if the symmetry-plane angles fluctuate with  $\eta$. 

At RHIC, the PHOBOS experiment reported the pseudorapidity dependence of elliptic flow over a wide range ($-5.0 < \eta < 5.3$) and variety of collision energies \cite{Back:2002gz,Back:2004mh,Back:2004zg}, and system sizes \cite{Alver:2006wh}. It was found that in the rest frame of one of the colliding nuclei ($\eta-y_{\text{beam}}$), $\mathrm{v}_2$ is energy independent. This feature was also observed in multiplicity density distributions \cite{Bearden:2001qq,Alver:2010ck} and for $\mathrm{v}_1$ \cite{Back:2005pc}. This suggests that at forward rapidity, in the fragmentation region, particle production is independent of the collision energy, an effect known as extended longitudinal scaling. 

In this Letter, we present measurements of $\mathrm{v}_2$, $\mathrm{v}_3$, and $\mathrm{v}_4$ over a wide pseudorapidity range ($-3.5 < \eta < 5.0$) in Pb--Pb collisions at $\sqrt{s_{\text{NN}}} = 2.76$ TeV using the ALICE detector. At the LHC, the pseudorapidity dependence of the flow harmonics has already been reported by ATLAS \cite{ATLAS:2012at,Aad:2014eoa} and CMS \cite{Chatrchyan:2012ta,Chatrchyan:2013kba} in a limited $\eta$-range ($|\eta| < 2.5$ and $|\eta| < 2.4$, respectively). The extended longitudinal scaling has been shown to hold for multiplicity densities \cite{Abbas:2013bpa} and directed flow \cite{Abelev:2013cva}, and appears to occur for elliptic flow \cite{Chatrchyan:2012ta,Aad:2014eoa}. Here, the $\eta$-range is extended considerably compared to the former results and we will investigate whether the extended longitudinal scaling of elliptic flow continues to hold. We will compare our data to hydrodynamical and transport models, and investigate the decrease of $\mathrm{v}_{n}$ in the forward regions relative to $\mathrm{d}N_{\mathrm{ch}}/\mathrm{d}\eta$.

\section{Experimental setup}
A detailed description of the ALICE detector is available elsewhere \cite{Aamodt:2008zz}. In this section, the sub-detectors used in this analysis are described: the V0 detector, the Time Projection Chamber (TPC), the Inner Tracking System (ITS) and the Forward Multiplicity Detector (FMD). The V0 detector consists of 2 arrays of scintillators located on opposite sides of the interaction point (IP) along the beam line. The detector has full azimuthal coverage in the ranges of $2.8 < \eta < 5.1$ (V0-A) and $-3.7 < \eta < -1.7$ (V0-C) \cite{Abbas:2013taa}. The detector acts as an online trigger and, with its large coverage, as a centrality estimator.

Charged particle tracks are reconstructed using the TPC, a large Time Projection Chamber \cite{Alme:2010ke}. The detector can provide position and momentum information. Particles that traverse the TPC volume leave ionization trails that drift towards the endcaps, where they are detected. Full length tracks can be reconstructed in the range $|\eta| < 0.8$. For this analysis, a transverse momentum range of $0.2 < p_{\text{T}} < 5.0$ GeV$/c$ was used. To ensure good track quality, the tracks are required to have at least 70 reconstructed TPC space points (cluster) out of 159 possible and an average $\chi^2$ per TPC cluster $\leq 4$. In addition, to reduce contamination from secondary particles (weak decays or interactions with material), a cut on the distance of closest approach (DCA) between the track and the primary vertex is applied both in the transverse plane (DCA$_{xy} < 2.4$ cm) and on the $z$-coordinate (DCA$_z < 3.2$ cm).

The ITS is made up of six cylindrical concentric silicon layers divided into three sub-systems, the Silicon Pixel Detector (SPD), the Silicon Drift Detector (SDD) and the Silicon Strip Detector (SSD), each consisting of two layers \cite{Aamodt:2008zz}. ITS clusters can be combined with the TPC information to improve track resolution. The SPD has additional applications \cite{Aamodt:2008zz}. Firstly, it is used to estimate the primary vertex as it is located close to the beam pipe. Secondly, clusters from the SPD inner layer, which consists of $3.3 \times 10^6$ pixels of size $50 \times 425$ $\mu$m$^2$, are used to estimate the number of charged particles in the range $|\eta| < 2.0$.

The FMD consists of five silicon rings, providing a pseudorapidity coverage in the ranges $-3.5 < \eta < -1.7$ and $1.7 < \eta < 5.0$ \cite{Christensen:2007yc}. The rings are single-layer detectors and only charged particle hits, not tracks, are measured. This means that primary and secondary particles cannot be distinguished. There are two types of FMD rings: inner ring and outer rings. Inner rings have 512 radial strips each covering $18^{\circ}$ in azimuth and outer rings have 256 radial strips each covering $9^{\circ}$ in azimuth. The charged particle estimation in the FMD is described in more detail elsewhere \cite{Abbas:2013bpa}. The inner layer of the SPD and the five FMD rings allow one to measure charged particle hits in the range $-3.5 < \eta < 5.0$.

\section{Data sample and analysis details}

We analysed $10$ million minimum bias Pb--Pb collisions at $\sqrt{s_{\text{NN}}}=2.76$ TeV. The sample was recorded during the first LHC heavy-ion data-taking period in 2010. A minimum-bias trigger requiring a coincidence between the signals from V0-A and V0-C was used. In addition, it is required that the primary vertex, determined by the SPD, be within $|\mathrm{v}_z| < 10.0$~cm, where $\mathrm{v}_{z} = 0$~cm is the location of the nominal interaction position. The measurements are grouped according to fractions of the inelastic cross section, and cover the $80\%$ most central collisions. The V0 detector is used for the centrality estimate which is described in more detail elsewhere \cite{Abelev:2013qoq}. For the most central to the most peripheral events, the V0 has a centrality resolution of $0.5\%$ to $2\%$, respectively.

The flow harmonics are estimated using the Q-cumulants method \cite{Bilandzic:2010jr} for two- and four-particle correlations, denoted as $\mathrm{v}_n\{2\}$ and $\mathrm{v}_n\{4\}$ respectively. The two- and four-particle cumulants respond differently to flow fluctuations. The two-particle cumulants are enhanced, while four-particle cumulants are suppressed. At forward rapidities, the pseudorapidity density is relatively low. This means that it is not always possible to get statistically significant results using only particles from a small region in $\eta$. To circumvent this using the Q-cumulants method, the reference flow measurement is performed using the charged particle tracks from the TPC, where the correlations at mid-rapidity are measured. As a systematic check, the charged particle tracks using a combination of the TPC and ITS are also used. Then, for the $v_n (\eta)$ analysis, the correlations between charged particle hits (from the SPD or FMD) and the tracks are measured in $\eta$-bins $0.5$ units of pseudorapidity wide. To avoid autocorrelations between the SPD clusters and tracks, the tracks for the reference particles are located in a different $\eta$-region than the SPD hits. Effectively, for SPD hits with $\eta < 0$, tracks are required to have $\eta > 0$ and vice versa. The same considerations apply for FMD hits. Due to the use of particle hits, only the $p_{\text{T}}$-integrated flow is measured. The $\phi$ distribution for the SPD or FMD clusters is not uniform, therefore a non-uniform acceptance correction is applied based on relations derived elsewhere \cite{Hansen:2014phd}.

As the inner rings of the FMD have only 20 azimuthal segments, the flow harmonics are slightly suppressed. The effect of this was recently calculated \cite{Bilandzic:2013kga} and found to be $1.6\%$, $3.7\%$ and $6.5\%$ for $\mathrm{v}_2$, $\mathrm{v}_3$ and $\mathrm{v}_4$ respectively. This suppression is taken into account in the final results. When using charged particle hits it is not possible to distinguish secondary particles (from material interactions and decays) from primary particles. For the regions covered by the SPD, the contamination from secondary particles is small ($< 10\%$), as the inner layer of the SPD is very close to the beam pipe. Away from mid-rapidity, in the FMD, dense material such as cooling tubes and read-out cables cause a very large production of secondary particles - up to twice the number of primary particles according to Monte Carlo (MC) studies. These secondary particles are deflected in $\varphi$ with respect to the mother particle, which causes a reduction in the observed flow. The reduction of flow caused by the secondary particles is estimated using an event generator containing particle yields, ratios, momentum spectra and flow coefficients, which are then subject to a full detector simulation using GEANT3 \cite{Brun:1994aa}. To make sure that the correction is not model dependent, the AMPT MC event generator \cite{Lin:2004en,Xu:2011fi} is used as an independent input, with GEANT3 again used to model the detector response. Using these simulations, the reduction is found to be larger for higher harmonics, up to $41\%$ for $\mathrm{v}_4$. Finally, the correction also accounts for missing very low $p_\text{T}$ particles, which increase the observed $v_{n}$ as these particles have a very small $v_n$. However, as the correction is always less than 1, the dominant effect comes from the secondary particles, which reduce $v_n$.

Few-particle correlations, not originating from the initial geometry termed non-flow (decays, jets, etc.), enhance the two-particle cumulant measurements. The non-flow contribution to the four-particle cumulant is found to be negligible \cite{Bilandzic:2010jr,Abelev:2014mda}, however, it is necessary to apply a correction to the two-particle cumulant. In the FMD and SPD, there is also a non-flow contribution from secondary particles, as they are sometimes produced in pairs. For the differential flow measurement, there is a rapidity-gap between the charged particle hits and the charged particle tracks. For the SPD, it is between 0 and 2 units in pseudorapidity, while for the FMD it is between 0.9 and 4.2 units in pseudorapidity. The large rapidity gap suppresses the non-flow contribution at forward rapidity. However, at mid-rapidities, this contribution is non-negligible and needs appropriate corrections. For the reference flow measurement there is no rapidity gap, and non-flow removal is important. For this analysis, the non-flow contributions are estimated using the HIJING event generator \cite{Wang:1991hta} and GEANT3 for the detector simulation. The non-flow contribution is estimated and subtracted separately for the reference and differential flow, before the correction for the deflection of secondary particles is applied and the $\mathrm{v}_n$ estimates are derived. 

\section{Systematic uncertainties}
Numerous sources of systematic uncertainty were investigated, including effects due to detector cuts, choice of reference particles and uncertainties related to the secondary particle correction. Four major contributors to the systematic uncertainty were identified: the choice of reference tracks, the model dependence of the secondary particle correction, the description of the detector used for the simulations, and finally the non-flow correction. As the non-flow contribution to the four-particle cumulant is negligible, only the first three systematic uncertainties are considered for $\mathrm{v}_2\{4\}$. The systematic uncertainties assigned to each of the sources are shown in Table~\ref{tab:sys}, and are described in more detail below.

\begin{table}[th!]
  \centering
  \begin{tabular}{l c c c c c}
    \hline
    \hline
    Source & $\mathrm{v}_2\{2\}$ & $\mathrm{v}_3\{2\}$ & $\mathrm{v}_4\{2\}$ & $\mathrm{v}_2\{4\}$ \\ 
    \hline
    Reference particle tracks & $2$-$4\%$ & $2$-$4\%$ & $2$-$6\%$ & $2$-$4\%$ \\
    Model dependence & $5\%$ & $5\%$ & $7\%$ & $5\%$ \\
    Material budget & $3$-$4\%$ & $3$-$4\%$ & $3$-$4\%$ & $3$-$4\%$ \\
    Non-flow correction & $2$-$10\%$ & $2$-$10\%$ & $2$-$10\%$ & - \\
    \hline
    Total & $6$-$12\%$ & $6$-$13\%$ & $6$-$14\%$ & $6$-$8\%$\\
    \hline
    \hline
  \end{tabular}
  \caption{List of the systematic uncertainties for each observable.}
  \label{tab:sys}
\end{table}
\begin{figure}[t!]
  \centering
  \includegraphics[width=\textwidth]{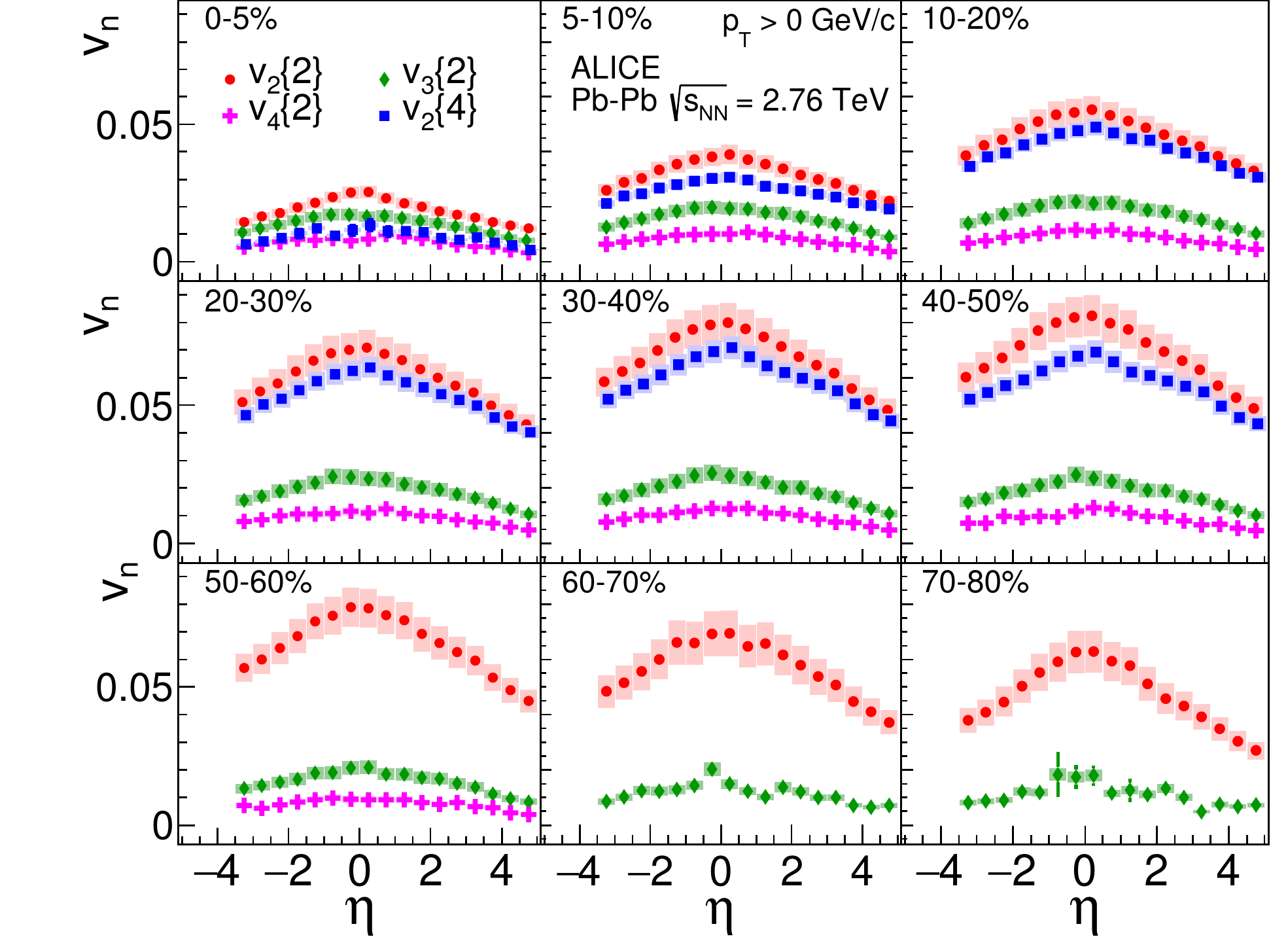}
  \caption{Measurements of the pseudorapidity dependence of $\mathrm{v}_2$, $\mathrm{v}_3$ and $\mathrm{v}_4$ in each centrality bin. The vertical lines represent the statistical uncertainties and the boxes represent the systematic uncertainties. The statistical uncertainties are usually smaller than the marker size.}
  \label{fig:panel}
\end{figure}
\begin{figure}[t!]
  \centering
  \includegraphics[width=0.7\textwidth]{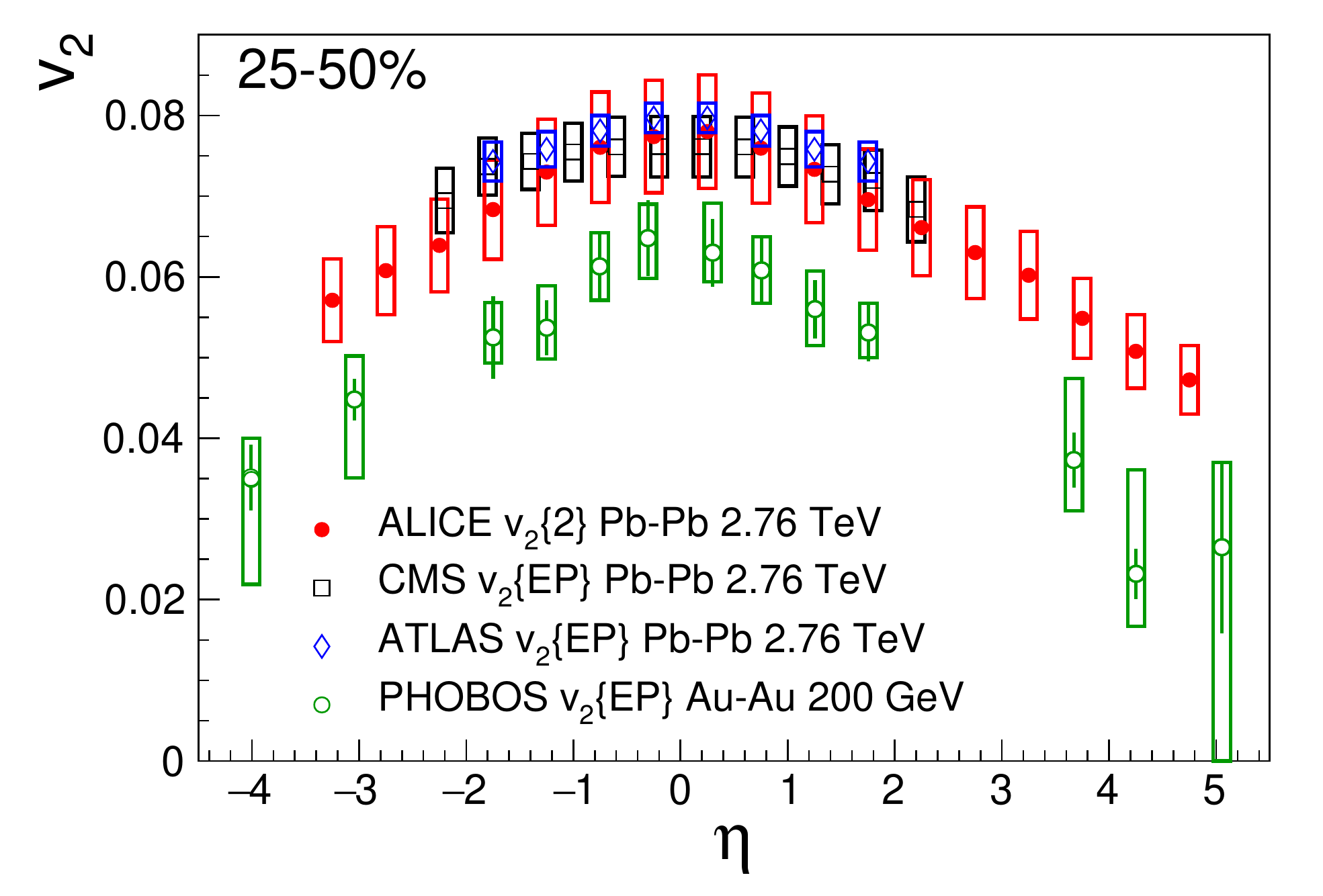}
  \caption{Elliptic flow for the $25$--$50\%$ centrality range. Boxes represent systematic uncertainties and errors bars represent statistical uncertainties. The results for $\mathrm{v}_2\{2\}$ from this analysis are compared to measurements using the event plane method from CMS \cite{Chatrchyan:2012ta} and ATLAS \cite{Aad:2014eoa} at the same energy and lower energy results from PHOBOS \cite{Back:2004mh}. For the comparable LHC energy, the $p_\text{T}$ range for ALICE is $p_\text{T} > 0 $ GeV/c, for CMS is $0.3 < p_\text{T} < 3 $ GeV/c, and for ATLAS is $p_\text{T} > 0.07 $ GeV/c.}
  \label{fig:rhic}
\end{figure}

The dependence of the differential flow on the reference tracks was tested by using tracks with combined information from the TPC and ITS, rather than tracks with only TPC information. The systematic uncertainty from the choice of reference tracks was found to vary slightly with centrality, with the most central events having the largest uncertainty. To test the model dependence of secondary particle production, the correction from the toy-model described above is compared to the one derived from AMPT tuned to LHC data. Both the secondary particle correction and the non-flow correction derived from HIJING are sensitive to inaccuracies in the description of the detector used for the simulation. To test this sensitivity, the output of two HIJING simulations with a flow afterburner, one with $+7\%$ material density and one with $-7\%$ material density, are compared to the output from having normal material density. In this case the systematic uncertainty has a small $\eta$-dependence, as there are significantly fewer secondary particles at mid-rapidity. The 3\% uncertainty is applicable to the SPD, while the 4\% uncertainty is applicable to the FMD.

We assessed the systematic uncertainty associated with the non-flow correction in two ways. Firstly, following another method proposed to subtract non-flow \cite{Voloshin:2008dg}, the two-particle cumulants were obtained from minimum bias pp collisions, where it is assumed that there is negligible anisotropic flow. The pp reference and differential cumulants are then rescaled according to their multiplicity, $M$, using the ratio $M^{\text{pp}}/M^{\text{cent}}$, then subtracted from the corresponding A-A cumulants. Any differences found between this method and the default HIJING method are treated as systematic uncertainties. Secondly, by using only charged particle hits from the SPD and FMD, it is possible to construct a two-particle cumulant with a large rapidity-gap, $\mathrm{v}_n\{2,|\Delta\eta|>2.0\}$, which largely removes all non-flow contributions. Unfortunately, this observable is  statistically stable only for $\mathrm{v}_2$ and $\mathrm{v}_3$, so it is used as a further cross check. In Table~\ref{tab:sys}, the 2\% uncertainties correspond to mid-central collisions where the ratio of flow to non-flow is largest, while the 10\% uncertainties correspond to very central and very peripheral collisions where the ratio of flow to non-flow is smallest. Finally, we used the AMPT model \cite{Lin:2004en,Xu:2011fi} to investigate if there are differences between $\mathrm{v}_{n}(\eta)$ and $\mathrm{v}_{n}(y)$, as $\eta$ is supposed to approximate $y$. We found there are 15\% differences in the flow coefficients at mid-rapidity, which reduced to 0\% for $\eta > 2$. We did not assign any systematic uncertainties due to these differences, as we are explicitly reporting measurements as a function of $\eta$ (as in the case of $\mathrm{d}N_{\mathrm{ch}}/\mathrm{d}\eta$ measurements).

The systematic uncertainty assigned to the non-flow correction is the largest contributor to the total systematic uncertainty, except for $\mathrm{v}_2\{4\}$ due to the four-particle cumulant's insensitivity to non-flow. The total systematic uncertainties are slightly dependent on centrality and pseudorapidity.

\begin{figure}[t!]
  \centering
  \includegraphics[width=0.7\textwidth]{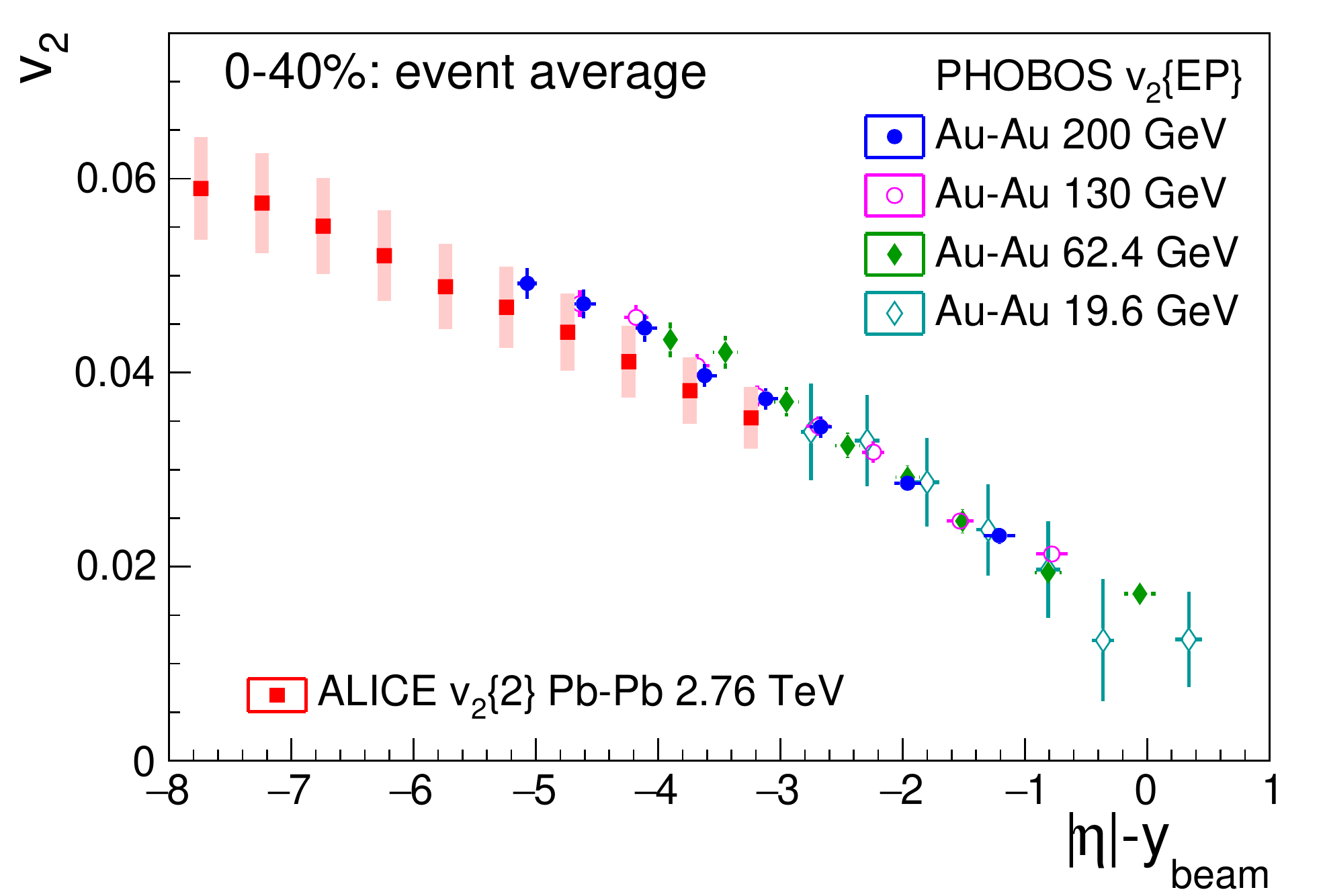}
  \caption{The elliptic flow as observed in the rest frame of one of the projectiles by using the variable $|\eta|-y_{beam}$ ($y_{beam}=7.99$) for the event averaged $0$--$40\%$ centrality range. The results from $\mathrm{v}_2\{2\}$ from this analysis are compared to lower energy results from PHOBOS \cite{Back:2004zg}. The vertical lines represent the statistical uncertainties and the boxes represent the systematic uncertainties. For the PHOBOS results only statistical errors are shown.}
  \label{fig:scaling}
\end{figure}
\begin{figure}[t!]
  \centering
  \includegraphics[width=0.7\textwidth]{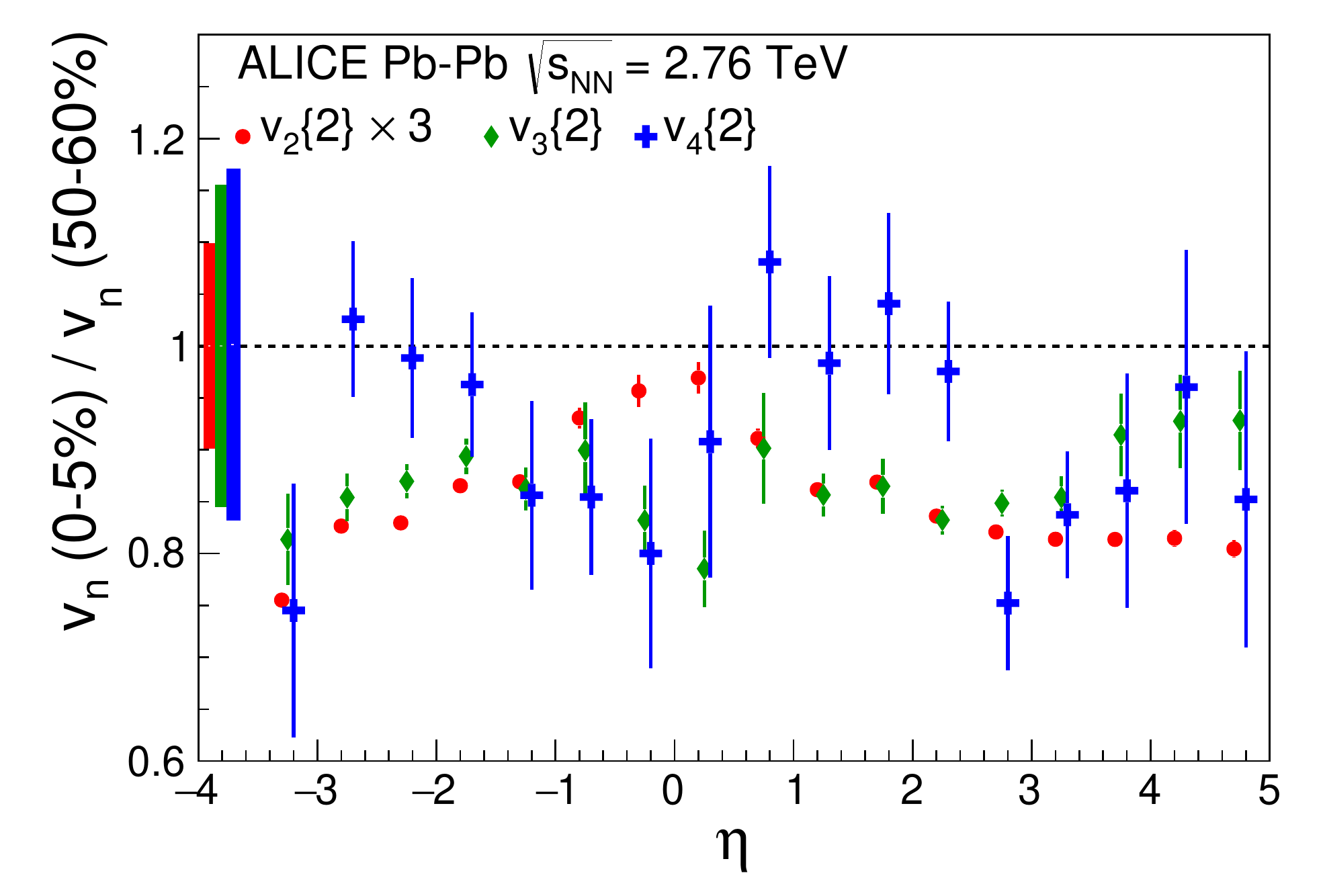}
  \caption{Ratio of $\mathrm{v}_n\{2\}$ between central ($0$--$5\%$) and peripheral ($50$--$60\%$) events for $\mathrm{v}_2$, $\mathrm{v}_3$ and $\mathrm{v}_4$. The vertical lines represent the statistical uncertainties and the boxes represent the systematic uncertainties. The $\mathrm{v}_2$ results are multiplied by 3 to fit on the same scale as $\mathrm{v}_3$ and $\mathrm{v}_4$.}
  \label{fig:cp}
\end{figure}

\begin{figure}[th!]
  \centering
  \includegraphics[width=0.7\textwidth]{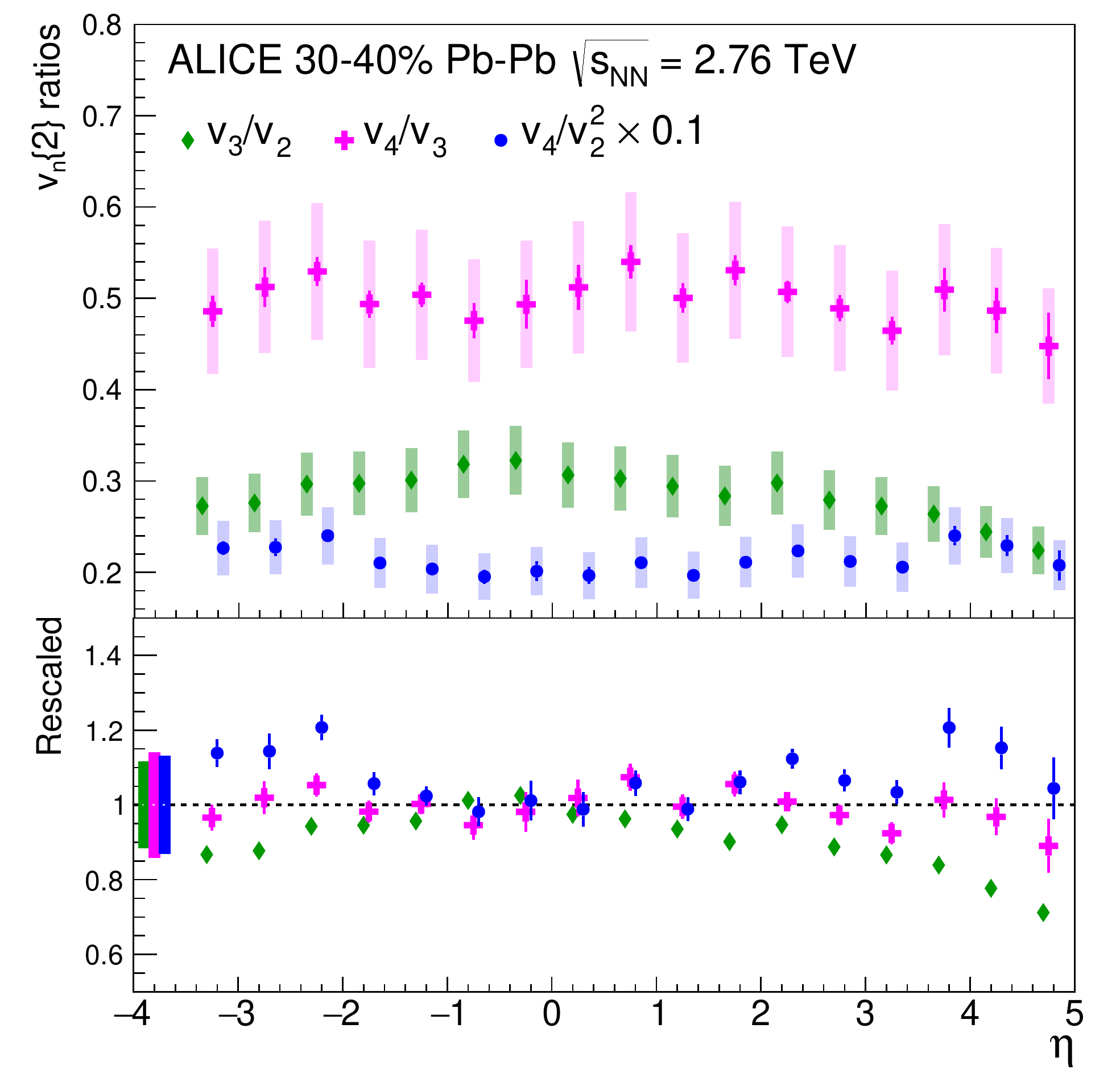}
  \caption{Ratios between different harmonics for the $30$--$40\%$ centrality range. The vertical lines represent the statistical uncertainties and the boxes represent the common systematic uncertainties. In the bottom panel the ratios are rescaled to $1$ at mid-rapidity and the common systematic uncertainties are shown as the thick bars on the left.}
  \label{fig:ratios}
\end{figure}

\section{Results}
An overview of the four observables in each centrality class is shown in Fig.~\ref{fig:panel}. Due to the changing overlap geometry, a strong centrality dependence of the elliptic flow is observed over the entire pseudorapidity range. The weaker centrality dependence of the higher order coefficients $\mathrm{v}_3$ and $\mathrm{v}_4$ is an indication that initial-state fluctuations play a prominent role, as the centrality dependence of the corresponding eccentricities are more modest relative to n=2 \cite{Alver:2010dn}. The different behaviour of $\mathrm{v}_2\{2\}$ and $\mathrm{v}_2\{4\}$ caused by flow fluctuations is also clearly seen. For the most peripheral events, there are not enough particles to get statistically stable results for $\mathrm{v}_2\{4\}$ and similarly for $\mathrm{v}_4\{2\}$ due to the relatively small quadrangular flow.

The $p_{\text{T}}$-integrated elliptic flow was also measured by CMS \cite{Chatrchyan:2012ta} and ATLAS \cite{Aad:2014eoa} in Pb--Pb collisions at $\sqrt{s_{\text{NN}}}=2.76$ TeV and by PHOBOS in Au--Au collisions at $\sqrt{s_{\text{NN}}}=200$ GeV \cite{Back:2004mh}. A comparison between those results and this analysis is shown for the $25$--$50\%$ centrality class in Fig.~\ref{fig:rhic}. In the common region of pseudorapidity acceptance, the results of present analysis are consistent with the results obtained by CMS and ATLAS experiments within the systematic uncertainties. The present analyses extends the measurements to a wider range of pseudorapidity. The values of $\mathrm{v}_{2}$ at all pseudorapidities measured at LHC energies are larger than the corresponding values at RHIC, as reported by PHOBOS. This increase in elliptic flow coincides with a larger $p_{\text{T}}$ at the LHC energy \cite{Aamodt:2010pa}.

The extended longitudinal scaling observed by PHOBOS in Au--Au collisions with center-of-mass energies from $19.6$ to $200$ GeV \cite{Back:2004zg} is found to hold up to the LHC energy (shown in Fig.~\ref{fig:scaling}). This is consistent with what was found by CMS \cite{Chatrchyan:2012ta} and ATLAS \cite{Aad:2014eoa}. Here it is shown as an event average for the $0$-$40\%$ most central events. The event average means that the analysis was performed in smaller centrality bins using multiplicity weights, and was then averaged over the centrality bins using the number of events as a weight \cite{Bilandzic:2010jr}. To examine boost invariance, it would be preferable to use rapidity ($y$) instead of pseudorapidity, unfortunately that is not possible using the FMD as the momentum cannot be measured. 

PHOBOS found the shape of $\mathrm{v}_2(\eta)$ to be largely independent of centrality, with only the overall level changing between central and peripheral events \cite{Back:2004mh}. The ratios of central to peripheral events for $\mathrm{v}_2$, $\mathrm{v}_3$ and $\mathrm{v}_4$ using the two-particle cumulant are shown in Fig.~\ref{fig:cp}. Here it is observed that none of the harmonics show a clear centrality dependence in the shape of $\mathrm{v}_n(\eta)$ within uncertainties (albeit hints of such a dependence are present in the $\mathrm{v}_2$ ratio), consistent with the results from PHOBOS at lower energy.

\begin{figure}[t]
  \centering
    \includegraphics[width=1\textwidth]{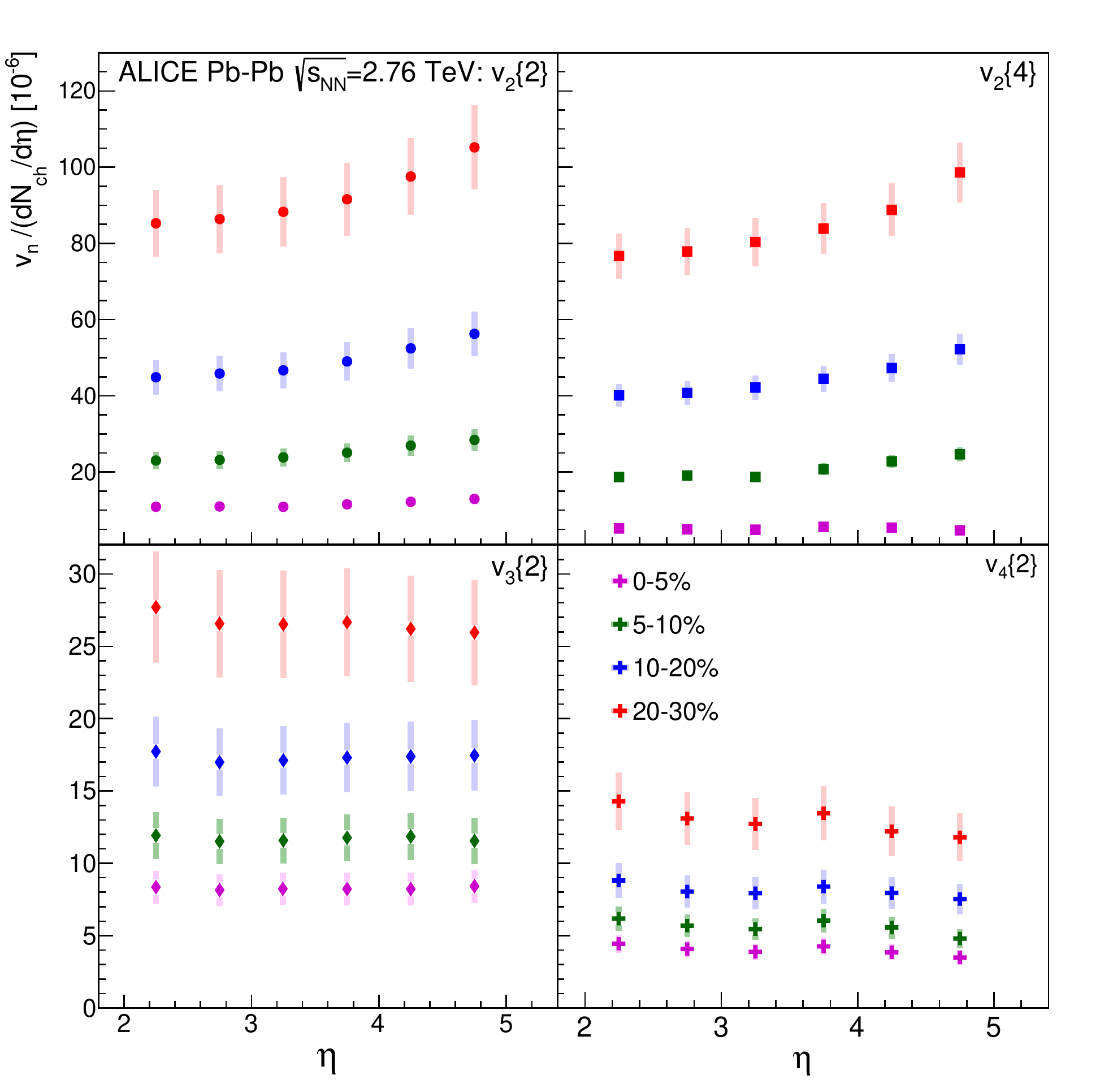}
  \caption{Ratios between $\mathrm{v}_{n}$ coefficients and $\mathrm{d}N_{\mathrm{ch}}/\mathrm{d}\eta$ values for different centralities. Measurements of $\mathrm{d}N_{\mathrm{ch}}/\mathrm{d}\eta$ are taken from a previous ALICE publication \cite{Abbas:2013bpa}. Only systematic uncertainties are shown, as the statistical uncertainties are smaller than the symbols.}
  \label{fig:ratiosdndeta}
\end{figure}

It is known that the suppression from viscous effects to the flow harmonics increases with $n$ \cite{Alver:2010dn}. The hadronic phase is speculated to be more dominant at forward rapidity \cite{Molnar:2014zha, Denicol:2015nhu}. Therefore, the relative decrease of the flow harmonics may help to disentangle the viscous effects from the hadronic phase with those from the QGP phase. When the ratio $\mathrm{v}_{m}/\mathrm{v}_{n}$ $(n\neq m)$ is formed most of the common systematic uncertainties cancel, leaving the contribution from the non-flow correction. The ratios of $\mathrm{v}_3/\mathrm{v}_2$ and $\mathrm{v}_4/\mathrm{v}_3$ are shown for the $30$-$40\%$ most central events in Fig.~\ref{fig:ratios}. A small decrease with $\eta$ is observed for $\mathrm{v}_3/\mathrm{v}_2$, qualitatively consistent with the expectation from viscous effects suppressing higher harmonics. The $\mathrm{v}_4/\mathrm{v}_3$ ratio remains constant with $|\eta|$ within the uncertainties. The figure also shows $\mathrm{v}_4/\mathrm{v}_2^2$, which is commonly used to estimate the non-linear contribution to $\mathrm{v}_4$ from the elliptic anisotropy \cite{Adams:2003zg}. Given the uncertainties, it is difficult to conclude whether $\mathrm{v}_4/\mathrm{v}_2^2$ changes with respect to $\eta$.

As mentioned previously, at forward rapidities the steepness of $\mathrm{v}_{n}(\eta)$ has been linked to the hadronic contribution to the viscosity to entropy ratio \cite{Molnar:2014zha, Denicol:2015nhu}. The larger the hadronic $\eta / s$, the steeper the fall off. We also note that the pseudorapidity densities of charged particles decrease in this region. In order to investigate the correspondence of the latter, in Fig.~\ref{fig:ratiosdndeta} we show the ratio of various $\mathrm{v}_n$ coefficients to previous ALICE measurements of $\mathrm{d}N_{\mathrm{ch}}/\mathrm{d}\eta$ \cite{Abbas:2013bpa}. In order to avoid any influence of the Jacobian translation from $y$ to $\eta$, only the range $\eta > 2$ is shown. We find that this ratio is generally flat, with the exception of $\mathrm{v}_{2}$ at the larger values of $\eta$. This indicates that within a fixed centrality interval, $\mathrm{v}_{3}$ and $\mathrm{v}_{4}$ are largely driven by the local particle density. Indeed, when comparing p--Pb and Pb--Pb collisions at LHC energies, it was found that values of $\mathrm{v}_3\{2\}$ were similar for similar values of $\mathrm{d}N_{\mathrm{ch}}/\mathrm{d}\eta$ \cite{Abelev:2014mda}. The correlation found between both quantities may be simply attributed to the fact that both particle production and the development of anisotropic flow are driven by the number of interactions in the system.

\begin{figure}
  \centering
  \includegraphics[width=0.7\textwidth]{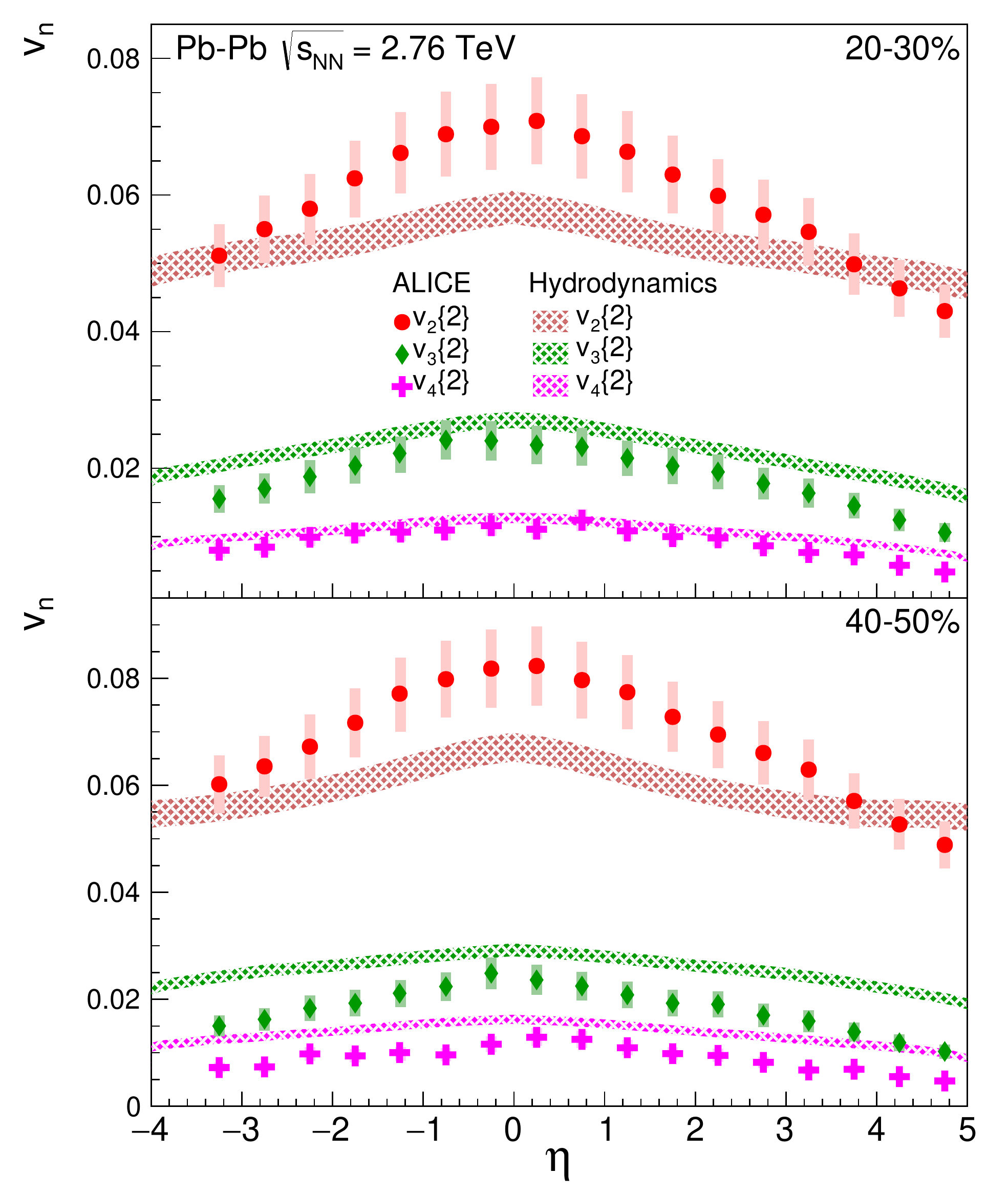}
  \caption{Comparisons to hydrodynamics predictions \cite{Denicol:2015nhu}, where input parameters (temperature dependence of  $\eta/s$) have been tuned to RHIC data for the Pb-Pb 20-30\% (top) and 40-50\% (bottom) centralities. The predictions are for Pb-Pb $\sqrt{s_{\text{NN}}} = 2.76$ TeV collisions. }
  \label{fig:hydro}
\end{figure}

In Fig.~\ref{fig:hydro}, we compare our data to hydrodynamic calculations tuned to RHIC data \cite{Denicol:2015nhu}. The tuning involves finding a parameterization of the temperature dependence of $\eta/s$, so that the hydrodynamical calculations describe PHOBOS measurements of $v_{2}(\eta)$ \cite{Back:2004mh,Back:2004zg}. It is clear that the same parameterization does not describe the LHC data as well. For both centralities, the elliptic flow coefficient $\mathrm{v}_2$ is generally underestimated, while the higher order coefficients $\mathrm{v}_3$ and  $\mathrm{v}_4$ are generally overestimated. This points to the need for an either an alternative parameterization of $\eta/s$ that describes both the RHIC and LHC data simultaneously, or further investigations into whether the initial state  model used is applicable for the LHC energies.

\begin{figure}
  \centering
  \includegraphics[width=0.7\textwidth]{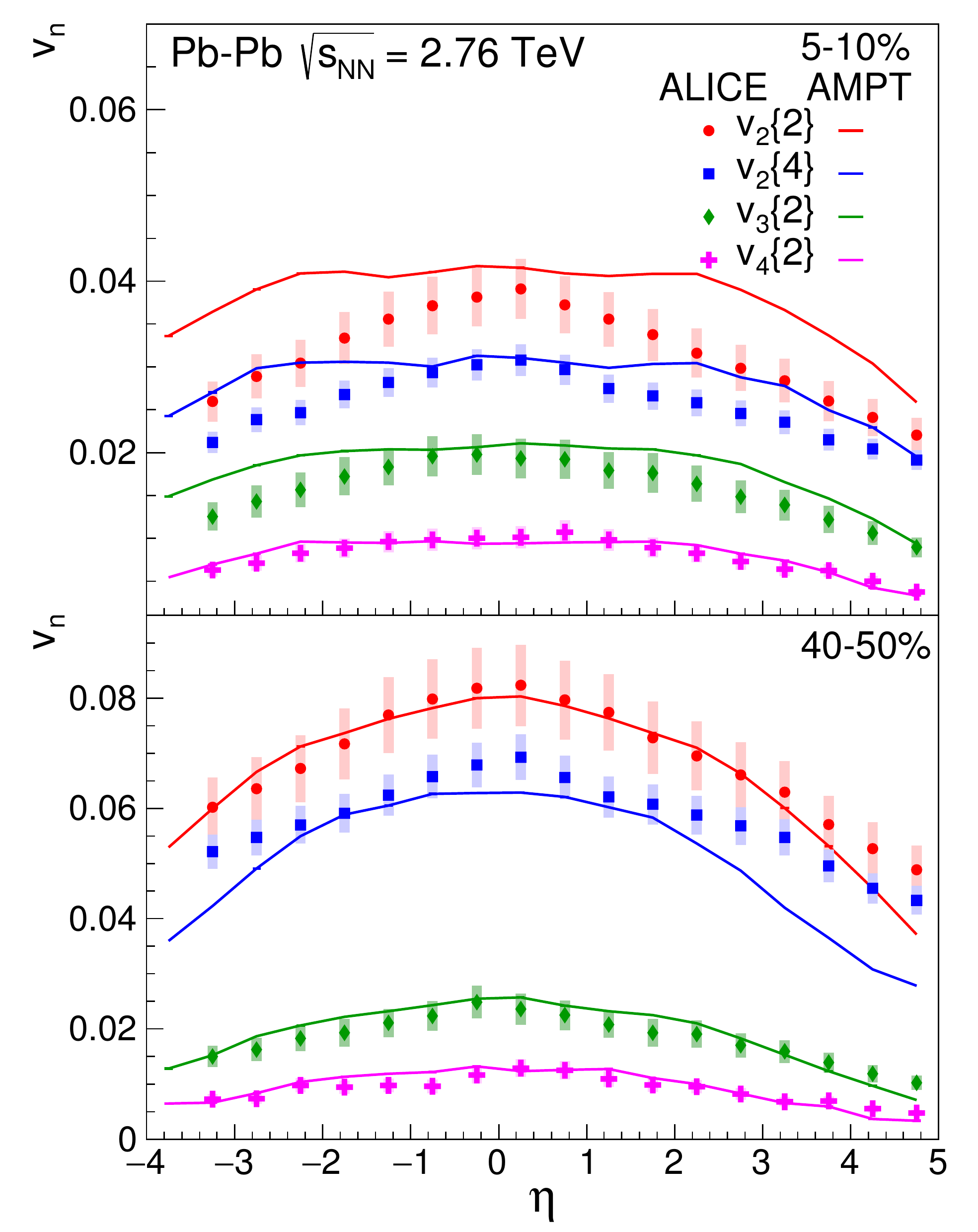}
  \caption{Comparison to AMPT \cite{Lin:2004en,Xu:2011fi} for the centrality ranges $5$--$10\%$ and (top) and $40$--$50\%$ (bottom). The AMPT predictions are for Pb-Pb $\sqrt{s_{\text{NN}}} = 2.76$ TeV collisions.}
  \label{fig:ampt}
\end{figure}

In contrast to hydrodynamical models,  AMPT is a non-equilibrium model that attempts to simulate parton production after the initial collision, and collective behaviour arises from parton and hadronic rescatterings. It has previously been tuned to agree with ALICE measurements of $\mathrm{v}_2$ vs. $p_{\text{T}}$ and multiplicity for the $40$-$50\%$ most central events. It was found to reproduce $\mathrm{v}_3(p_{\text{T}})$ well using the same parameters. In Fig.~\ref{fig:ampt} the results of this analysis are compared to the output of the AMPT model for two different centralities. For the centrality range of $40$--$50\%$, which AMPT is tuned to match, there is good agreement at mid-rapidity for all observables modulo $\mathrm{v}_2\{4\}$ at larger $|\eta|$, where AMPT underestimates the data. The underestimation at forward rapidity is found to be independent of the choice of reference particles, suggesting that it is unrelated to symmetry plane angle fluctuations with $\eta$. For more central events AMPT tends to overestimate flow at forward rapidities, except for $\mathrm{v}_4$ which it describes quite well over the entire range. At mid-rapidity AMPT agrees with the observed values of $\mathrm{v}_2$, $\mathrm{v}_3$ and $\mathrm{v}_4$ within the systematic uncertainties. Further tuning may lead to an improvement at forward rapidities, and should be investigated in future studies.

\section{Conclusions}
The pseudorapidity dependence of the anisotropic flow harmonics $\mathrm{v}_2$, $\mathrm{v}_3$ and $\mathrm{v}_4$ have been measured in Pb--Pb collisions at $\sqrt{s_{\text{NN}}}=2.76$ TeV using the ALICE detector. The measurement is performed over the widest $\eta$-range at the LHC, $-3.5 < \eta < 5.0$, in nine centrality bins covering $0$ to $80\%$ of the total inelastic cross section. It was found that the shape of $\mathrm{v}_n(\eta)$ does not depend obviously on centrality. Comparing to lower energy measurements at RHIC, elliptic flow is larger at the LHC over the entire pseudorapidity range and extended longitudinal scaling of $\mathrm{v}_2$ observed at lower collision energies is still valid up to the LHC energy. In the range $|\eta| < 2.5$ the results were found to be consistent with previous LHC measurements. At forward rapidities, the higher harmonic flow coefficients are proportional to the charged particle densities for a given centrality, while the ratio of $\mathrm{v}_2$ to $\mathrm{d}N_{\mathrm{ch}}/\mathrm{d}\eta$ rises with increasing $\eta$. A comparison to hydrodynamic calculations tuned to RHIC data has difficulties in describing our data in some $\eta$ regions, and this suggests that the LHC data play a key role in constraining either the temperature dependence of $\eta/s$ or the initial state. Finally, comparing our data to AMPT, the model describes the flow well at mid-rapidity, but fails for $\mathrm{v}_2$ at forward rapidities. 
%

\newenvironment{acknowledgement}{\relax}{\relax}
\begin{acknowledgement}
\section*{Acknowledgements}

The ALICE Collaboration would like to thank all its engineers and technicians for their invaluable contributions to the construction of the experiment and the CERN accelerator teams for the outstanding performance of the LHC complex.
The ALICE Collaboration gratefully acknowledges the resources and support provided by all Grid centres and the Worldwide LHC Computing Grid (WLCG) collaboration.
The ALICE Collaboration acknowledges the following funding agencies for their support in building and
running the ALICE detector:
State Committee of Science,  World Federation of Scientists (WFS)
and Swiss Fonds Kidagan, Armenia;
Conselho Nacional de Desenvolvimento Cient\'{\i}fico e Tecnol\'{o}gico (CNPq), Financiadora de Estudos e Projetos (FINEP),
Funda\c{c}\~{a}o de Amparo \`{a} Pesquisa do Estado de S\~{a}o Paulo (FAPESP);
Ministry of Science \& Technology of China (MSTC), National Natural Science Foundation of China (NSFC) and Ministry of Education of China (MOEC)";
Ministry of Science, Education and Sports of Croatia and  Unity through Knowledge Fund, Croatia;
Ministry of Education and Youth of the Czech Republic;
Danish Natural Science Research Council, the Carlsberg Foundation and the Danish National Research Foundation;
The European Research Council under the European Community's Seventh Framework Programme;
Helsinki Institute of Physics and the Academy of Finland;
French CNRS-IN2P3, the `Region Pays de Loire', `Region Alsace', `Region Auvergne' and CEA, France;
German Bundesministerium fur Bildung, Wissenschaft, Forschung und Technologie (BMBF) and the Helmholtz Association;
General Secretariat for Research and Technology, Ministry of Development, Greece;
National Research, Development and Innovation Office (NKFIH), Hungary;
Council of Scientific and Industrial Research (CSIR), New Delhi;
Department of Atomic Energy and Department of Science and Technology of the Government of India;
Istituto Nazionale di Fisica Nucleare (INFN) and Centro Fermi - Museo Storico della Fisica e Centro Studi e Ricerche ``Enrico Fermi'', Italy;
Japan Society for the Promotion of Science (JSPS) KAKENHI and MEXT, Japan;
National Research Foundation of Korea (NRF);
Consejo Nacional de Cienca y Tecnologia (CONACYT), Direccion General de Asuntos del Personal Academico(DGAPA), M\'{e}xico, Amerique Latine Formation academique - 
European Commission~(ALFA-EC) and the EPLANET Program~(European Particle Physics Latin American Network);
Stichting voor Fundamenteel Onderzoek der Materie (FOM) and the Nederlandse Organisatie voor Wetenschappelijk Onderzoek (NWO), Netherlands;
Research Council of Norway (NFR);
Pontificia Universidad Cat\'{o}lica del Per\'{u};
National Science Centre, Poland;
Ministry of National Education/Institute for Atomic Physics and National Council of Scientific Research in Higher Education~(CNCSI-UEFISCDI), Romania;
Joint Institute for Nuclear Research, Dubna;
Ministry of Education and Science of Russian Federation, Russian Academy of Sciences, Russian Federal Agency of Atomic Energy, Russian Federal Agency for Science and Innovations and The Russian Foundation for Basic Research;
Ministry of Education of Slovakia;
Department of Science and Technology, South Africa;
Centro de Investigaciones Energeticas, Medioambientales y Tecnologicas (CIEMAT), E-Infrastructure shared between Europe and Latin America (EELA), 
Ministerio de Econom\'{i}a y Competitividad (MINECO) of Spain, Xunta de Galicia (Conseller\'{\i}a de Educaci\'{o}n),
Centro de Aplicaciones Tecnológicas y Desarrollo Nuclear (CEA\-DEN), Cubaenerg\'{\i}a, Cuba, and IAEA (International Atomic Energy Agency);
Swedish Research Council (VR) and Knut $\&$ Alice Wallenberg Foundation (KAW);
National Science and Technology Development Agency (NSDTA), Suranaree University of Technology (SUT) and Office of the Higher Education Commission under NRU project of Thailand;
Ukraine Ministry of Education and Science;
United Kingdom Science and Technology Facilities Council (STFC);
The United States Department of Energy, the United States National Science Foundation, the State of Texas, and the State of Ohio.

\end{acknowledgement}

\bibliographystyle{utphys}   
\bibliography{bibliography}
\newpage
%
\appendix
\section{The ALICE Collaboration}
\label{app:collab}



\begingroup
\small
\begin{flushleft}
J.~Adam\Irefn{org39}\And
D.~Adamov\'{a}\Irefn{org85}\And
M.M.~Aggarwal\Irefn{org89}\And
G.~Aglieri Rinella\Irefn{org35}\And
M.~Agnello\Irefn{org111}\And
N.~Agrawal\Irefn{org48}\And
Z.~Ahammed\Irefn{org134}\And
S.~Ahmad\Irefn{org19}\And
S.U.~Ahn\Irefn{org69}\And
S.~Aiola\Irefn{org138}
% ...
\renewcommand\labelenumi{\textsuperscript{\theenumi}~}

\section*{Affiliation notes}
\renewcommand\theenumi{\roman{enumi}}
\begin{Authlist}
\item \Adef{0}Deceased
\item \Adef{idp1793936}{Also at: Georgia State University, Atlanta, Georgia, United States}
\item \Adef{idp3203424}{Also at: Also at Department of Applied Physics, Aligarh Muslim University, Aligarh, India}
\item \Adef{idp3917824}{Also at: M.V. Lomonosov Moscow State University, D.V. Skobeltsyn Institute of Nuclear, Physics, Moscow, Russia}
\end{Authlist}

\section*{Collaboration Institutes}
\renewcommand\theenumi{\arabic{enumi}~}
\begin{Authlist}

\item \Idef{org1}A.I. Alikhanyan National Science Laboratory (Yerevan Physics Institute) Foundation, Yerevan, Armenia
\item \Idef{org2}Benem\'{e}rita Universidad Aut\'{o}noma de Puebla, Puebla, Mexico
\item \Idef{org3}Bogolyubov Institute for Theoretical Physics, Kiev, Ukraine
\item \Idef{org4}Bose Institute, Department of Physics and Centre for Astroparticle Physics and Space Science (CAPSS), Kolkata, India
\item \Idef{org5}Budker Institute for Nuclear Physics, Novosibirsk, Russia
\item \Idef{org6}California Polytechnic State University, San Luis Obispo, California, United States
\item \Idef{org7}Central China Normal University, Wuhan, China
\item \Idef{org8}Centre de Calcul de l'IN2P3, Villeurbanne, France
\item \Idef{org9}Centro de Aplicaciones Tecnol\'{o}gicas y Desarrollo Nuclear (CEADEN), Havana, Cuba
\item \Idef{org10}Centro de Investigaciones Energ\'{e}ticas Medioambientales y Tecnol\'{o}gicas (CIEMAT), Madrid, Spain
\item \Idef{org11}Centro de Investigaci\'{o}n y de Estudios Avanzados (CINVESTAV), Mexico City and M\'{e}rida, Mexico
\item \Idef{org12}Centro Fermi - Museo Storico della Fisica e Centro Studi e Ricerche ``Enrico Fermi'', Rome, Italy
\item \Idef{org13}Chicago State University, Chicago, Illinois, USA
\item \Idef{org14}China Institute of Atomic Energy, Beijing, China
\item \Idef{org15}Commissariat \`{a} l'Energie Atomique, IRFU, Saclay, France
\item \Idef{org16}COMSATS Institute of Information Technology (CIIT), Islamabad, Pakistan
\item \Idef{org17}Departamento de F\'{\i}sica de Part\'{\i}culas and IGFAE, Universidad de Santiago de Compostela, Santiago de Compostela, Spain
\item \Idef{org18}Department of Physics and Technology, University of Bergen, Bergen, Norway
\item \Idef{org19}Department of Physics, Aligarh Muslim University, Aligarh, India
\item \Idef{org20}Department of Physics, Ohio State University, Columbus, Ohio, United States
\item \Idef{org21}Department of Physics, Sejong University, Seoul, South Korea
\item \Idef{org22}Department of Physics, University of Oslo, Oslo, Norway
\item \Idef{org23}Dipartimento di Fisica dell'Universit\`{a} 'La Sapienza' and Sezione INFN Rome, Italy
\item \Idef{org24}Dipartimento di Fisica dell'Universit\`{a} and Sezione INFN, Cagliari, Italy
\item \Idef{org25}Dipartimento di Fisica dell'Universit\`{a} and Sezione INFN, Trieste, Italy
\item \Idef{org26}Dipartimento di Fisica dell'Universit\`{a} and Sezione INFN, Turin, Italy
\item \Idef{org27}Dipartimento di Fisica e Astronomia dell'Universit\`{a} and Sezione INFN, Bologna, Italy
\item \Idef{org28}Dipartimento di Fisica e Astronomia dell'Universit\`{a} and Sezione INFN, Catania, Italy
\item \Idef{org29}Dipartimento di Fisica e Astronomia dell'Universit\`{a} and Sezione INFN, Padova, Italy
\item \Idef{org30}Dipartimento di Fisica `E.R.~Caianiello' dell'Universit\`{a} and Gruppo Collegato INFN, Salerno, Italy
\item \Idef{org31}Dipartimento di Scienze e Innovazione Tecnologica dell'Universit\`{a} del  Piemonte Orientale and Gruppo Collegato INFN, Alessandria, Italy
\item \Idef{org32}Dipartimento Interateneo di Fisica `M.~Merlin' and Sezione INFN, Bari, Italy
\item \Idef{org33}Division of Experimental High Energy Physics, University of Lund, Lund, Sweden
\item \Idef{org34}Eberhard Karls Universit\"{a}t T\"{u}bingen, T\"{u}bingen, Germany
\item \Idef{org35}European Organization for Nuclear Research (CERN), Geneva, Switzerland
\item \Idef{org36}Excellence Cluster Universe, Technische Universit\"{a}t M\"{u}nchen, Munich, Germany
\item \Idef{org37}Faculty of Engineering, Bergen University College, Bergen, Norway
\item \Idef{org38}Faculty of Mathematics, Physics and Informatics, Comenius University, Bratislava, Slovakia
\item \Idef{org39}Faculty of Nuclear Sciences and Physical Engineering, Czech Technical University in Prague, Prague, Czech Republic
\item \Idef{org40}Faculty of Science, P.J.~\v{S}af\'{a}rik University, Ko\v{s}ice, Slovakia
\item \Idef{org41}Faculty of Technology, Buskerud and Vestfold University College, Vestfold, Norway
\item \Idef{org42}Frankfurt Institute for Advanced Studies, Johann Wolfgang Goethe-Universit\"{a}t Frankfurt, Frankfurt, Germany
\item \Idef{org43}Gangneung-Wonju National University, Gangneung, South Korea
\item \Idef{org44}Gauhati University, Department of Physics, Guwahati, India
\item \Idef{org45}Helmholtz-Institut f\"{u}r Strahlen- und Kernphysik, Rheinische Friedrich-Wilhelms-Universit\"{a}t Bonn, Bonn, Germany
\item \Idef{org46}Helsinki Institute of Physics (HIP), Helsinki, Finland
\item \Idef{org47}Hiroshima University, Hiroshima, Japan
\item \Idef{org48}Indian Institute of Technology Bombay (IIT), Mumbai, India
\item \Idef{org49}Indian Institute of Technology Indore, Indore (IITI), India
\item \Idef{org50}Indonesian Institute of Sciences, Jakarta, Indonesia
\item \Idef{org51}Inha University, Incheon, South Korea
\item \Idef{org52}Institut de Physique Nucl\'eaire d'Orsay (IPNO), Universit\'e Paris-Sud, CNRS-IN2P3, Orsay, France
\item \Idef{org53}Institut f\"{u}r Informatik, Johann Wolfgang Goethe-Universit\"{a}t Frankfurt, Frankfurt, Germany
\item \Idef{org54}Institut f\"{u}r Kernphysik, Johann Wolfgang Goethe-Universit\"{a}t Frankfurt, Frankfurt, Germany
\item \Idef{org55}Institut f\"{u}r Kernphysik, Westf\"{a}lische Wilhelms-Universit\"{a}t M\"{u}nster, M\"{u}nster, Germany
\item \Idef{org56}Institut Pluridisciplinaire Hubert Curien (IPHC), Universit\'{e} de Strasbourg, CNRS-IN2P3, Strasbourg, France
\item \Idef{org57}Institute for Nuclear Research, Academy of Sciences, Moscow, Russia
\item \Idef{org58}Institute for Subatomic Physics of Utrecht University, Utrecht, Netherlands
\item \Idef{org59}Institute for Theoretical and Experimental Physics, Moscow, Russia
\item \Idef{org60}Institute of Experimental Physics, Slovak Academy of Sciences, Ko\v{s}ice, Slovakia
\item \Idef{org61}Institute of Physics, Academy of Sciences of the Czech Republic, Prague, Czech Republic
\item \Idef{org62}Institute of Physics, Bhubaneswar, India
\item \Idef{org63}Institute of Space Science (ISS), Bucharest, Romania
\item \Idef{org64}Instituto de Ciencias Nucleares, Universidad Nacional Aut\'{o}noma de M\'{e}xico, Mexico City, Mexico
\item \Idef{org65}Instituto de F\'{\i}sica, Universidad Nacional Aut\'{o}noma de M\'{e}xico, Mexico City, Mexico
\item \Idef{org66}iThemba LABS, National Research Foundation, Somerset West, South Africa
\item \Idef{org67}Joint Institute for Nuclear Research (JINR), Dubna, Russia
\item \Idef{org68}Konkuk University, Seoul, South Korea
\item \Idef{org69}Korea Institute of Science and Technology Information, Daejeon, South Korea
\item \Idef{org70}KTO Karatay University, Konya, Turkey
\item \Idef{org71}Laboratoire de Physique Corpusculaire (LPC), Clermont Universit\'{e}, Universit\'{e} Blaise Pascal, CNRS--IN2P3, Clermont-Ferrand, France
\item \Idef{org72}Laboratoire de Physique Subatomique et de Cosmologie, Universit\'{e} Grenoble-Alpes, CNRS-IN2P3, Grenoble, France
\item \Idef{org73}Laboratori Nazionali di Frascati, INFN, Frascati, Italy
\item \Idef{org74}Laboratori Nazionali di Legnaro, INFN, Legnaro, Italy
\item \Idef{org75}Lawrence Berkeley National Laboratory, Berkeley, California, United States
\item \Idef{org76}Moscow Engineering Physics Institute, Moscow, Russia
\item \Idef{org77}Nagasaki Institute of Applied Science, Nagasaki, Japan
\item \Idef{org78}National Centre for Nuclear Studies, Warsaw, Poland
\item \Idef{org79}National Institute for Physics and Nuclear Engineering, Bucharest, Romania
\item \Idef{org80}National Institute of Science Education and Research, Bhubaneswar, India
\item \Idef{org81}National Research Centre Kurchatov Institute, Moscow, Russia
\item \Idef{org82}Niels Bohr Institute, University of Copenhagen, Copenhagen, Denmark
\item \Idef{org83}Nikhef, Nationaal instituut voor subatomaire fysica, Amsterdam, Netherlands
\item \Idef{org84}Nuclear Physics Group, STFC Daresbury Laboratory, Daresbury, United Kingdom
\item \Idef{org85}Nuclear Physics Institute, Academy of Sciences of the Czech Republic, \v{R}e\v{z} u Prahy, Czech Republic
\item \Idef{org86}Oak Ridge National Laboratory, Oak Ridge, Tennessee, United States
\item \Idef{org87}Petersburg Nuclear Physics Institute, Gatchina, Russia
\item \Idef{org88}Physics Department, Creighton University, Omaha, Nebraska, United States
\item \Idef{org89}Physics Department, Panjab University, Chandigarh, India
\item \Idef{org90}Physics Department, University of Athens, Athens, Greece
\item \Idef{org91}Physics Department, University of Cape Town, Cape Town, South Africa
\item \Idef{org92}Physics Department, University of Jammu, Jammu, India
\item \Idef{org93}Physics Department, University of Rajasthan, Jaipur, India
\item \Idef{org94}Physik Department, Technische Universit\"{a}t M\"{u}nchen, Munich, Germany
\item \Idef{org95}Physikalisches Institut, Ruprecht-Karls-Universit\"{a}t Heidelberg, Heidelberg, Germany
\item \Idef{org96}Purdue University, West Lafayette, Indiana, United States
\item \Idef{org97}Pusan National University, Pusan, South Korea
\item \Idef{org98}Research Division and ExtreMe Matter Institute EMMI, GSI Helmholtzzentrum f\"ur Schwerionenforschung, Darmstadt, Germany
\item \Idef{org99}Rudjer Bo\v{s}kovi\'{c} Institute, Zagreb, Croatia
\item \Idef{org100}Russian Federal Nuclear Center (VNIIEF), Sarov, Russia
\item \Idef{org101}Saha Institute of Nuclear Physics, Kolkata, India
\item \Idef{org102}School of Physics and Astronomy, University of Birmingham, Birmingham, United Kingdom
\item \Idef{org103}Secci\'{o}n F\'{\i}sica, Departamento de Ciencias, Pontificia Universidad Cat\'{o}lica del Per\'{u}, Lima, Peru
\item \Idef{org104}Sezione INFN, Bari, Italy
\item \Idef{org105}Sezione INFN, Bologna, Italy
\item \Idef{org106}Sezione INFN, Cagliari, Italy
\item \Idef{org107}Sezione INFN, Catania, Italy
\item \Idef{org108}Sezione INFN, Padova, Italy
\item \Idef{org109}Sezione INFN, Rome, Italy
\item \Idef{org110}Sezione INFN, Trieste, Italy
\item \Idef{org111}Sezione INFN, Turin, Italy
\item \Idef{org112}SSC IHEP of NRC Kurchatov institute, Protvino, Russia
\item \Idef{org113}Stefan Meyer Institut f\"{u}r Subatomare Physik (SMI), Vienna, Austria
\item \Idef{org114}SUBATECH, Ecole des Mines de Nantes, Universit\'{e} de Nantes, CNRS-IN2P3, Nantes, France
\item \Idef{org115}Suranaree University of Technology, Nakhon Ratchasima, Thailand
\item \Idef{org116}Technical University of Ko\v{s}ice, Ko\v{s}ice, Slovakia
\item \Idef{org117}Technical University of Split FESB, Split, Croatia
\item \Idef{org118}The Henryk Niewodniczanski Institute of Nuclear Physics, Polish Academy of Sciences, Cracow, Poland
\item \Idef{org119}The University of Texas at Austin, Physics Department, Austin, Texas, USA
\item \Idef{org120}Universidad Aut\'{o}noma de Sinaloa, Culiac\'{a}n, Mexico
\item \Idef{org121}Universidade de S\~{a}o Paulo (USP), S\~{a}o Paulo, Brazil
\item \Idef{org122}Universidade Estadual de Campinas (UNICAMP), Campinas, Brazil
\item \Idef{org123}University of Houston, Houston, Texas, United States
\item \Idef{org124}University of Jyv\"{a}skyl\"{a}, Jyv\"{a}skyl\"{a}, Finland
\item \Idef{org125}University of Liverpool, Liverpool, United Kingdom
\item \Idef{org126}University of Tennessee, Knoxville, Tennessee, United States
\item \Idef{org127}University of the Witwatersrand, Johannesburg, South Africa
\item \Idef{org128}University of Tokyo, Tokyo, Japan
\item \Idef{org129}University of Tsukuba, Tsukuba, Japan
\item \Idef{org130}University of Zagreb, Zagreb, Croatia
\item \Idef{org131}Universit\'{e} de Lyon, Universit\'{e} Lyon 1, CNRS/IN2P3, IPN-Lyon, Villeurbanne, France
\item \Idef{org132}Universit\`{a} di Brescia
\item \Idef{org133}V.~Fock Institute for Physics, St. Petersburg State University, St. Petersburg, Russia
\item \Idef{org134}Variable Energy Cyclotron Centre, Kolkata, India
\item \Idef{org135}Warsaw University of Technology, Warsaw, Poland
\item \Idef{org136}Wayne State University, Detroit, Michigan, United States
\item \Idef{org137}Wigner Research Centre for Physics, Hungarian Academy of Sciences, Budapest, Hungary
\item \Idef{org138}Yale University, New Haven, Connecticut, United States
\item \Idef{org139}Yonsei University, Seoul, South Korea
\item \Idef{org140}Zentrum f\"{u}r Technologietransfer und Telekommunikation (ZTT), Fachhochschule Worms, Worms, Germany
\end{Authlist}
\endgroup

\end{document}